\documentclass[sigconf]{acmart}
\AtBeginDocument{%
  }

\usepackage[inline]{enumitem}
\usepackage{subcaption} 
\usepackage{float}

\copyrightyear{2025}
\acmYear{2025}
\setcopyright{rightsretained}
\acmConference[DLfM 2025]{12th International Conference on Digital Libraries for Musicology}{September 26, 2025}{Seoul, Republic of Korea}
\acmBooktitle{12th International Conference on Digital Libraries for Musicology (DLfM 2025), September 26, 2025, Seoul, Republic of Korea}
\acmPrice{}
\acmDOI{10.1145/3748336.3748340}
\acmISBN{979-8-4007-2083-3/25/09}

\begin{document}

\title{Drafting the Landscape of Computational Musicology Tools: a Survey-Based Approach}


\author{Jorge Morgado-Vega, Sachin Sharma, Federico Simonetta}
\affiliation{%
  \institution{GSSI -- Gran Sasso Science Institute}
  \city{L'Aquila}
  \country{Italy}
}



\renewcommand{\shortauthors}{Morgado et al.}

\begin{abstract}
Since the 60s, musicology has been increasingly impacted by computational tools in various ways, from systematic analysis approaches to modeling of creativity. This article presents a comprehensive assessment of the current state of Computational Musicology tools based on survey data collected from practitioners in the field. We gathered information on tool usage patterns, common analytical tasks, user satisfaction levels, data characteristics, and prioritized features across four distinct domains: symbolic music, music-related imagery, audio, and text. Our findings reveal significant gaps between current tooling capabilities and user needs, highlighting some limitations of these tools across all domains. This assessment contributes to the ongoing dialogue between tool developers and music scholars, aiming to enhance the effectiveness and accessibility of computational methods in musicological research.
\end{abstract}




\maketitle

\section{Introduction}\label{sec:intro}

\begin{figure*}[htbp]
    \begin{subfigure}[m]{0.49\textwidth}
        \centering
        \includegraphics[width=\linewidth]{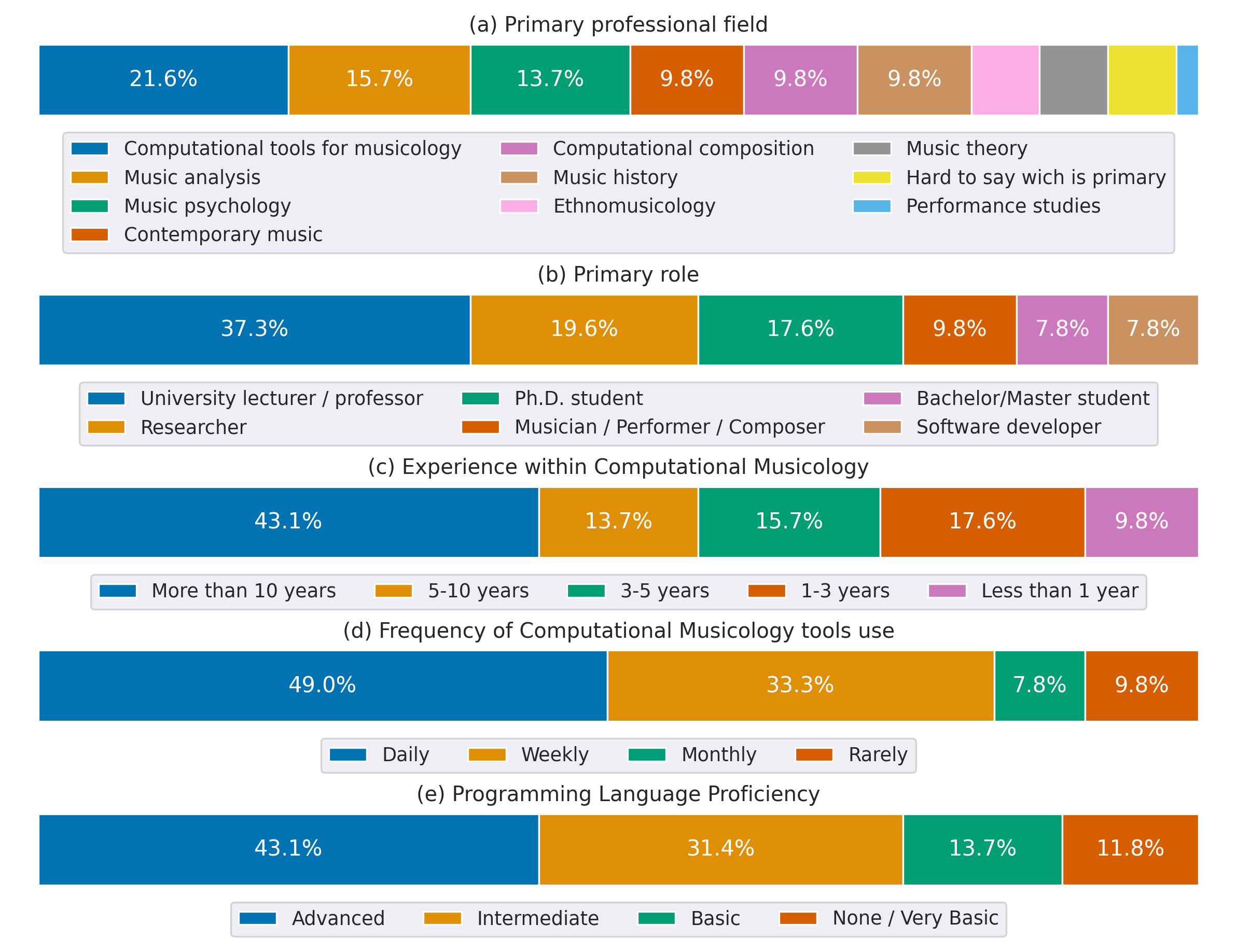}
        \caption{Professional experience of the participants}
        \label{fig:demographics}
    \end{subfigure}
    \begin{subfigure}[m]{0.49\textwidth}
        \centering
        \includegraphics[width=0.75\linewidth]{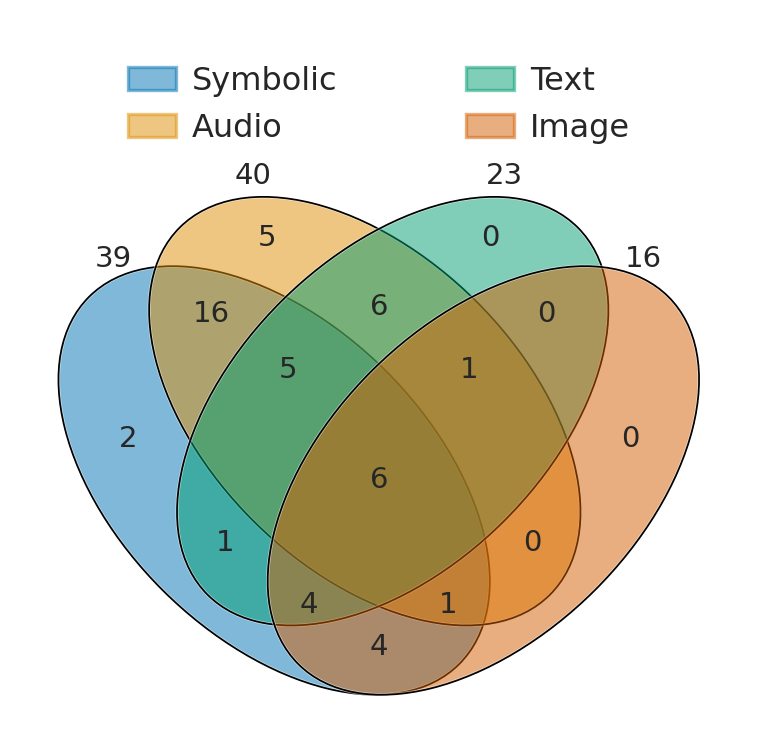}
        \caption{Amount of respondents who reported to work with each combination of modalities.}
        \label{fig:venn}
    \end{subfigure}
    \caption{Description of the survey participants}
\end{figure*}

Computational approaches in musicological study have been documented since the 1960s~\cite{berlind1966addendum,lincoln1970current}. These pioneering efforts occurred during the mainframe period, when computing capabilities were limited and highly specialized. The subsequent proliferation of hardware, expansion of programming language options, joint to the advent of personal computing technology, facilitated the development of increasingly specialized analytical processes, enabling individual researchers to conduct Computational Musicology (CM) projects autonomously~\cite{hewlett1991computing} and leading CM to gain importance in the whole context of musicology~\cite{Cook2004}.

Today's landscape offers a diverse array of tools spanning multiple domains of computational music analysis research. These tools extend beyond basic statistical score analysis to encompass audio manipulation, algorithmic composition, data retrieval, and numerous other applications.

The multimodal nature of music~\cite{simonetta2019multimodal} has contributed to the creation of heterogeneous manipulation approaches and software tools. These divergent implementation paradigms create substantial interoperability challenges, often requiring researchers to convert between multiple data formats or master several different interfaces depending on the task.

This technical fragmentation exacerbates the communication gap between musicologists and software developers, which has been well documented since the first years of development~\cite{larue1970some}. Despite decades of CM development, these communities frequently operate in parallel with limited mutual understanding~\cite{cheng2017educational}. Musicologists may struggle to articulate technical requirements in terms familiar to developers, while developers may create powerful tools that fail to address musicologists' actual research questions or workflow patterns.

In this article, we use a survey-based approach to investigate the main needs and concerns of CM practitioners, thus helping strengthen the connections between user requirements and available technologies in the future. The main contributions of this work are:
\begin{enumerate}
    \item An attempt to understand the state of CM tools from a perspective as objective as possible, in order to inform the future development of tools.
    \item Quantitative results supporting the identification of strengths and weaknesses of the existing CM tools.
    \item A discussion of potential approaches to limit the problems that emerged from this work.
\end{enumerate}

The paper is organized as follows: Section~\ref{sec:related-works} summarizes previous research in this area; Section~\ref{sec:methodology} details the survey methodology applied to musicology practitioners; Section \ref{sec:results} presents the findings from the survey implementation; Section~\ref{sec:discussion} offers our interpretation of the data, highlights the main issues that exist in the CM field, and proposes approaches to tackle them; Section~\ref{sec:conclusions} offers concluding observations.

\section{Related works}\label{sec:related-works}

The use of technology in musicology has been widely studied in recent decades. Inskip and Wiering~\cite{inskip2015their} conducted a survey of musicologists, examining their primary specializations, digital skills, and views on the risks and limitations of using technology in musicological research. Additional studies~\cite{panteli2018review,volk2012mathematical} have explored both the application of computational tools in musicology and the challenges that arise when combining different disciplines such as mathematics, computer science, and musicology. Specifically, Volk and Honingh~\cite{volk2012mathematical} highlighted the lack of a comprehensive approach due to the development of many different models for specific aspects of music.

Reflecting this inherently multimodal nature of musical phenomena, CM utilizes a diverse range of tools tailored to different representations, or modalities, of music~\cite{simonetta2019multimodal}. As computing resources expanded, a broader array of tools emerged, each addressing specific modalities and research objectives.

For symbolic music data, such as scores encoded in formats like Kern, MusicXML, or MIDI, a variety of tools facilitate computational analysis and feature extraction. Early tools like the command-line interface (CLI) based Humdrum Toolkit~\cite{huron2002music} focused on processing structured text representations of scores. Python libraries such as music21~\cite{cuthbert2010music21} provide extensive capabilities for parsing, manipulating, and analyzing symbolic structures. Tools like jSymbolic~\cite{mckay2018jsymbolic} and the more recent Musif~\cite{llorens2023musif,simonetta2023optimizing} specialize in extracting high-level features from symbolic formats, commonly used for tasks like style, genre, or composer identification, a subfield that has been systematically surveyed by \citet{simonetta2025stylebased}. Emerging approaches also explore graph-based representations and tools like graphmuse~\cite{karystinaios2024graphmuse} for complex structural analysis.

Complementing symbolic analysis, tools for processing audio data address the acoustic dimension of music. Libraries such as librosa~\cite{mcfee2015librosa}, Essentia~\cite{bogdanov2013essentia}, and Torchaudio~\cite{yang2022torchaudio} provide foundational capabilities for signal processing, feature extraction, and deep learning on audio signals, essential for tasks in Music Information Retrieval (MIR). To support research reproducibility, mirdata~\cite{bittner2019mirdata} standardizes access to commonly used MIR datasets.

Beyond programmatic libraries, graphical user interfaces (GUIs) and interactive environments enhance accessibility and support visual or real-time workflows. Applications like Sonic Visualiser~\cite{rector2018sonic} offer visual analysis of audio features (e.g., spectrograms), while OpenMusic~\cite{agon1998openmusic} provides a visual environment for algorithmic composition. PureData~\cite{puckette1996pure} and Max/MSP are widely used for interactive and real-time music generation and performance in experimental and artistic contexts. These tools also exemplify different interface paradigms: some offer graphical user interfaces (e.g., OpenMusic~\cite{agon1998openmusic}, Sonic Visualiser~\cite{rector2018sonic}, Puredata~\cite{puckette1996pure}), prioritizing visual feedback, while others employ command-line interfaces (e.g., Humdrum Toolkit~\cite{huron2002music}) optimized for automation. Still others exist primarily as programming libraries (e.g., music21~\cite{cuthbert2010music21}), requiring integration into codebases. Some tools even offer multiple interfaces (e.g., jSymbolic~\cite{mckay2018jsymbolic}).

Despite this rich ecosystem of tools spanning various modalities, the field faces persistent challenges in usability, interoperability, and alignment with musicological perspectives. Usability remains a significant barrier, particularly for musicologists without extensive programming backgrounds, as noted by Ariza~\cite{Ariza2008}, while developers may lack deep musicological insight. Furthermore, evaluating computational models solely based on technical metrics like classification accuracy may overlook musicologically meaningful interpretations, a concern raised by Sturm~\cite{Sturm2014}. Bridging the gap between computational features and human perception is also vital~\cite{simonetta2022perceptual}. Pearce and Wiggins~\cite{PearceWiggins2012} emphasized the necessity of incorporating cognitive and perceptual models to support theoretically grounded research in music cognition and analysis.

This non-exhaustive summary of CM tools highlights a heterogeneous landscape that sets problematic aspects in interoperability, usability, and alignment with musicological perspectives. However, how these issues relate to the single tools, modalities, and musicological tasks remains obscure. In this article, we try to shed light on the limits of the existing CM tools.

\begin{figure*}[htbp]
    \centering
    \includegraphics[width=\linewidth]{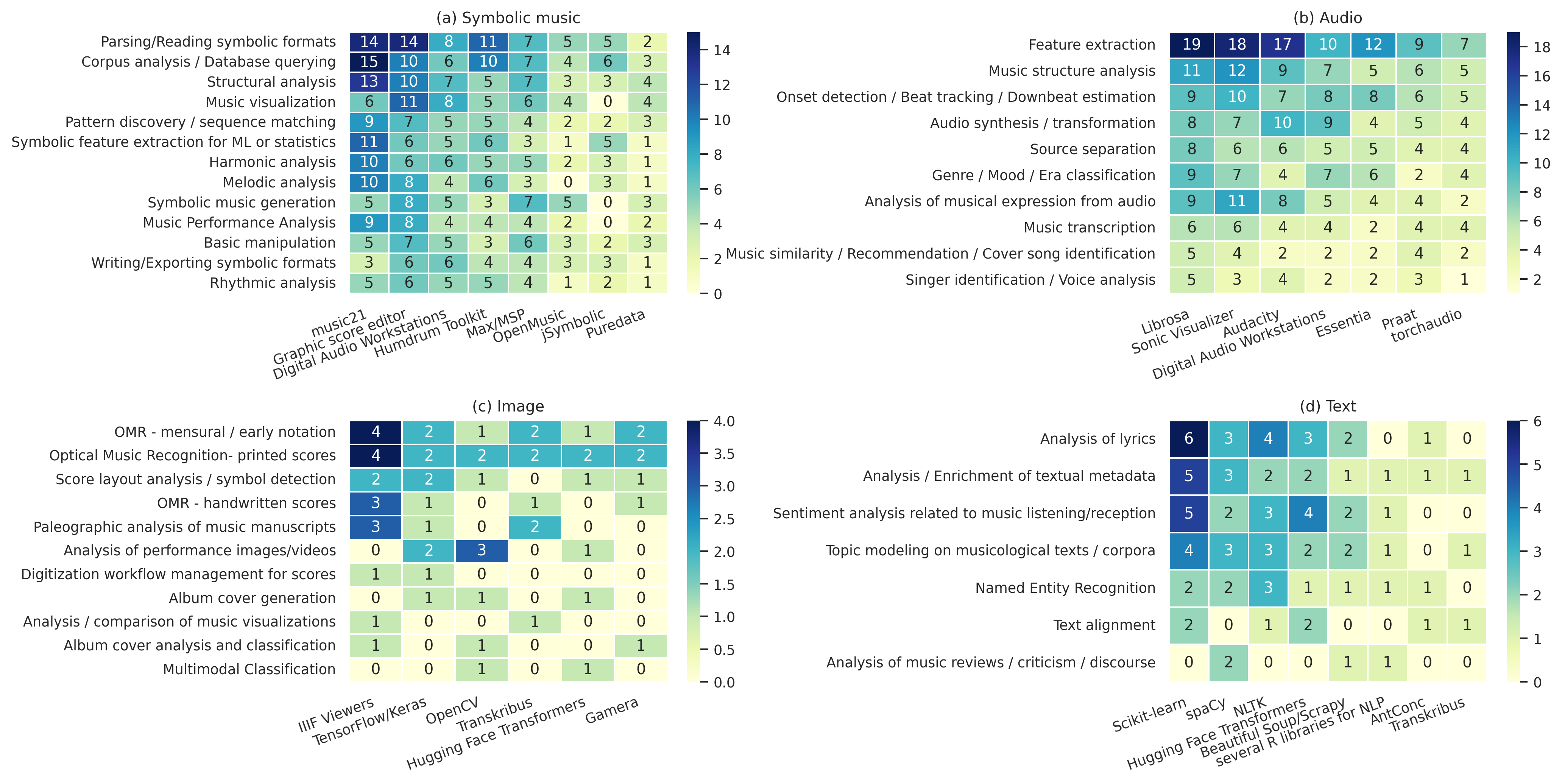}
    \caption{Tasks vs. tools reported. Each value represents the number of participants who reported performing a task X (among all that they perform) and also reported using a tool Y (among all they use). Values are filtered to show the top 80\% common matches within each domain.}
    \label{fig:tools-vs-tasks}
\end{figure*}

\section{Methodology}\label{sec:methodology}

To assess the current state of CM tools, we conducted a comprehensive online survey targeting practitioners in the field (e.g., Professors, Researchers, PhD. Students, Software Developers). The survey was designed to collect quantitative and qualitative data on tool usage patterns, user satisfaction, and unmet needs in four primary modalities of music data:
\begin{enumerate*}
    \item symbolic music,
    \item audio music,
    \item music-related images, and
    \item lyrics and music-related texts.
\end{enumerate*}

The questionnaire was distributed across authors' direct contacts and public academic mailing lists, including international communities (ISMIR), italian-based music computing associations (AIMI), and UK/US-based mailing lists about musicology (JISCMail). It must be noted that although these communities are mainly Europe and North America-centric, they still include a range of international professionals. 

The subsequent sections provide a detailed overview of each part of the survey.

\subsection{Participant demographics}

The first questions were dedicated to knowing the background of each interviewee.

\textbf{What is your primary role in relation to CM?} This question helps us understand the professional distribution of the interviewees and enables analysis of how different roles might influence tool preferences and research approaches. Possible values were: Student (Bachelor/Master); PhD Student; Post-doctoral researcher; Researcher; University Lecturer/Professor; Software Developer; Music Archivist / Librarian / Digital Curator; Musician / Performer / Composer using computational tools; Other (with the possibility of writing them). Upon analysis, in order to avoid confusion, we decided to merge the classes "Researcher" and "Post-doctoral researcher".

\textbf{What is your primary field of research within musicology?} By identifying participants' specific research domains, we can assess how computational methods are being applied across different musicological subdisciplines. Possible values were: Music History; Contemporary music; Ethnomusicology; Music theory; Performance studies; Algorithmic/computational composition; Music psychology; Music neurology; Computational tools for musicology (R\&D); Music analysis; Popular music; Other (With the possibility of writing them).

\textbf{For how many years have you been actively working or studying in the field of CM?} Understanding experience levels allows us to distinguish between early-career and more established participants, which can impact perspectives on computational methods. Possible values were: Less than 1 year; 1-3 years; 3-5 years; 5-10 years; More than 10 years.

\textbf{What is your level of proficiency in programming and using software libraries?} This assesses technical fluency, which is crucial for interpreting how users engage with computational musicology tools. Possible values were: None / Very Basic (e.g., only using GUIs; Basic: Using existing scripts/tools, perhaps with minor modifications; Intermediate: Writing/modifying complex scripts, using library APIs, combining tools; Advanced: Developing complex libraries/tools, contributing to open-source projects.

\textbf{How often do you use computational tools for your research?} The frequency of use gives insight into how integral computational tools are to the respondent's research practice. Possible values were: Daily, Weekly, Monthly, and Rarely.

\subsection{Modality-specific inquiries}

\begin{figure*}[htbp]
    \centering
    \includegraphics[width=\linewidth]{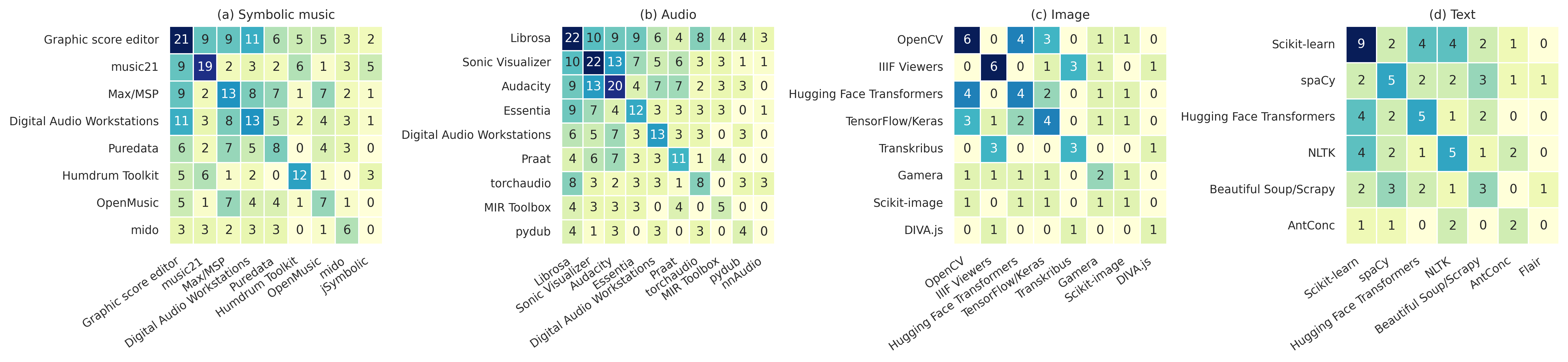}
    \caption{Tools vs. tools reported. Each value represents the number of participants who reported using a pair of tools X and Y together (among all they use). Values are filtered to show the top 80\% common matches within each domain.}
    \label{fig:tools-vs-tools}
\end{figure*}
We then asked participants whether they have experience working in each of the four modalities. Based on their responses, for each modality in which they indicated involvement, we presented a series of modality-specific questions.

\textbf{Which specific tasks are the most common in your workflow?} This question helps us understand the practical activities carried out within each modality and identify common workflows. We provided a list of common tasks tailored to each modality. Participants could select as many tasks as they regularly perform, and also had the option to specify additional tasks not listed.

\textbf{Name the libraries, software tools, or platforms that you use the most frequently for the mentioned tasks? (at most 5)} This allows us to identify which tools are most widely adopted and assess their prevalence across different research contexts. A curated list of tools was presented for each modality. Participants could also add tools not included in the list.

\textbf{How satisfied are you, overall, with the tools/libraries that you mentioned?} Understanding satisfaction levels helps us evaluate whether existing tools meet users’ needs and expectations. Participants rated their satisfaction on a scale from 1 to 5, with 5 indicating "Completely satisfied" and 1 meaning "Not satisfied at all".

\textbf{What do you not like about them?} This question aims to highlight common usability issues and barriers to effective tool adoption. Possible values were: Difficult installation; Hard to learn; Has major bugs; Poor documentation; Slow performance; Difficulty in integration with other tools; Limited functionality; Other (with the possibility of writing it).

\textbf{How large are the corpora you typically analyze?} Corpora size informs us about the scale of computational work and potential performance or scalability requirements. Possible values were: Tiny collections (1-10 pieces); Small collections (10-100 pieces); Medium collection (100-1000 pieces); Large collections (1000-1M pieces); Massive collections (1M+ pieces).

\textbf{List tasks for which you lack adequate tools. What specific features are missing or need improvement?} This helps identify gaps in current toolsets and user needs that are not being met by available solutions. Participants could freely describe tasks for which they feel existing tools are insufficient, and detail the features they believe are missing or need enhancement.

\subsection{Final questions}

We also aimed to understand participants' broader practices, preferences, and challenges when working within the field of Computational Musicology. These questions address how users find resources, choose tools, and manage data integration across modalities.

\textbf{Where do you usually look for support, new libraries, or information within the field of Computational Musicology?} This question aims to map the main channels through which users access support and stay informed, highlighting community practices and resource visibility. Possible values where: Official documentation of libraries/tools; GitHub (Issues, Discussions, Code search); Stack Overflow; Specific mailing lists (e.g., ISMIR community, music-ir, audiodsp, specific library lists); Academic conferences (e.g., ISMIR, ICMC, CMMR, DLfM, TENOR, SMC) \& proceedings; Specific forums (e.g., KVR, ModWiggler, specific software forums); Academic papers / Theses; Blogs / Online tutorials; Colleagues / Personal contacts; Other (with the possibility to write them).

\textbf{How important are the following factors when choosing a computational musicology/MIR library?
} Participants were asked to rate the importance on a scale from 1 to 5 (with 5 being the highest importance) of the following factors: Ease of integration with other software; Availability of well-documented API and examples; Speed and efficiency of processing; Cross-platform compatibility (e.g., Windows, MacOS, Linux); Active community support; Availability of pre-trained models; Easy-to-use graphical interface; Programming Language. This helps identify the criteria that guide tool selection, offering insight into user priorities and expectations when adopting new libraries. 

\textbf{Do you encounter difficulties integrating data, methods, or tools from the different data types you worked with (e.g., aligning audio and symbolic data, linking text analysis to musical events, combining OMR output with audio analysis)?} This question assesses the technical and methodological challenges of multimodal integration, a critical aspect in many computational musicology projects. Possible values were: Yes, No, or Not Applicable. Participants who reported difficulties were invited to describe their experiences.

\section{Results}\label{sec:results}

We collected a total of 59 responses from participants representing diverse research fields and professional roles. From these, we removed answers that lacked at least one modality and one tool, resulting in 51 final answers. The following sections present the significant findings that emerged from the survey analysis.

Figure~\ref{fig:demographics} details respondent demographics and expertise. Primary research areas (a) included mainly musicological fields, with computational-related topics summing to 35.9\%. It must be noted that, during the collection of responses, the category "Hard to say" was named "Other" and had space for users' custom description; it has been successively renamed upon analysis of the answers. As expected, most users were academics, with 37.3\% being lecturers or professors, making up nearly 75\% of those involved in research (b). Regarding CM tool experience, most were highly experienced (>70\% with $\ge$3 years, and >40\% with $\ge$10 years) and used tools daily or weekly (c, d). Programming proficiency varied (e); a majority reported advanced or intermediate skills, with others having basic or no experience.

As shown in Figure~\ref{fig:venn}, symbolic music analysis and audio analysis stood as the most common modalities among practitioners.

Figure~\ref{fig:tools-vs-tasks}a reveals how both graphical user interface tools (e.g., graphic score editors, Digital Audio Workstations) and non-GUI tools (e.g., music21, Humdrum toolkit) are notably employed for tasks dealing with the symbolic music modality. Moreover, visual programming language tools (Max/MSP, PureData, OpenMusic) are also used remarkably frequently, presumably for music generation. 

On the audio modality side, Figure~\ref{fig:tools-vs-tasks}b shows feature extraction as the most common task, followed by onset detection / beat tracking / downbeat estimation, and music structure analysis. The most used tools were reported to be Sonic Visualiser and the Python library \textit{librosa}, further reinforcing how both graphical user interface and non-GUI tools are widely used across different domains.

Unsurprisingly, the most common tasks within the image modality are related to musical scores and Optical Music Recognition (OMR) -- see Figure~\ref{fig:tools-vs-tasks}c. Related tools are IIIF viewers for Digital Humanities collections and Deep Learning tools (Tensorflow/Keras). Regarding texts, Figure~\ref{fig:tools-vs-tasks}d shows the prevalence of Python libraries for analyzing lyrics and textual metadata.

Analyzing the co-occurrences of the tools, it is clear that GUIs are often used in conjunction by the same users using non-GUIs software -- see Figure~\ref{fig:tools-vs-tools}. For instance, graphic score editors were used by 9 people together with music21, while librosa was used by 10 people aside from Sonic Visualizer and by 9 people in conjunction with Audacity. This suggests that in the audio and symbolic modalities GUI software represents a complementary solution and not an alternative to non-GUI frameworks. This is partially confirmed by Figure~\ref{fig:tools-vs-tasks}.

\begin{figure}[tb]
    \centering
    \includegraphics[width=\linewidth]{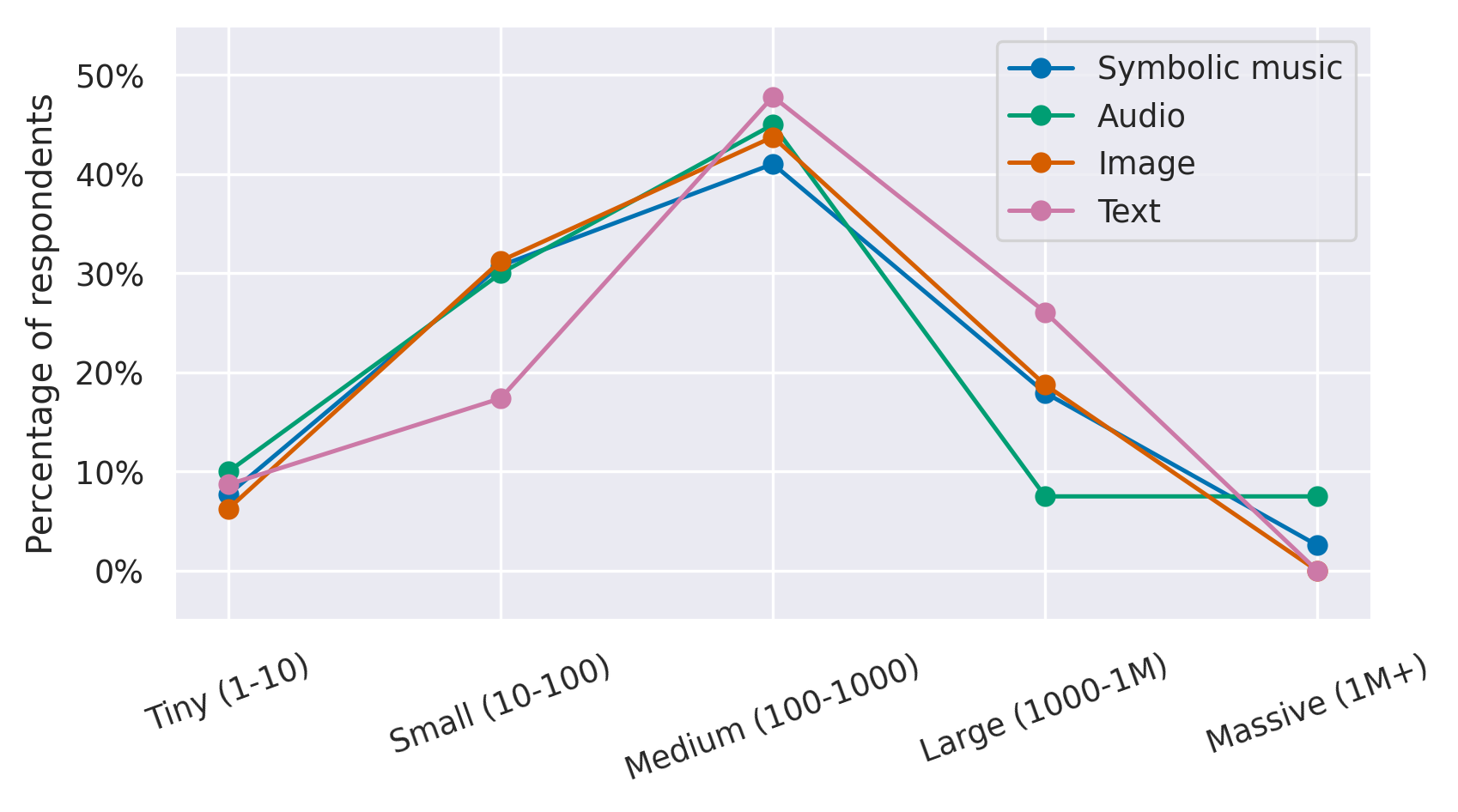}
    \caption{Corpora size}
    \label{fig:corpora-size}
\end{figure}

When asked about the typical data size, medium-sized collections (100-1,000 items) dominate research practices, representing more than 40\% of the cases on each modality -- see Figure~\ref{fig:corpora-size}. 

\begin{figure}[htbp]
    \centering
    \includegraphics[width=\linewidth]{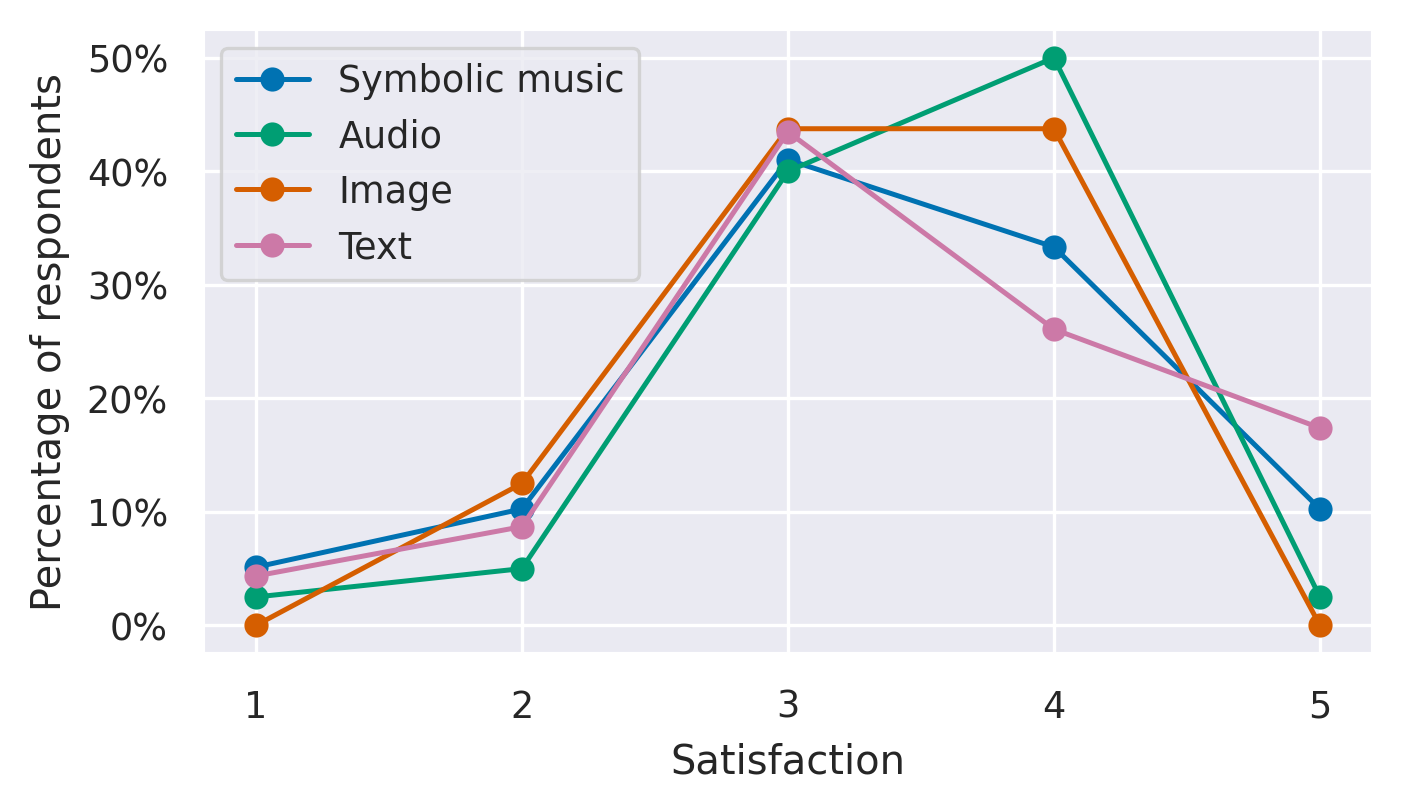}
    \caption{Percentage of answers across satisfaction levels per modality. 1="Not satisfied at all", 5="Completely satisfied"}
    \label{fig:all-sat}
\end{figure}

Across all four modalities, satisfaction levels were predominantly in the
neutral to satisfied range -- see Figure~\ref{fig:all-sat}.
The average satisfaction level was 3.33 for symbolic music, 3.45 for audio, 3.31
for images, and 3.43 for text.

\begin{figure}[tb]
    \centering
    \includegraphics[width=.8\linewidth]{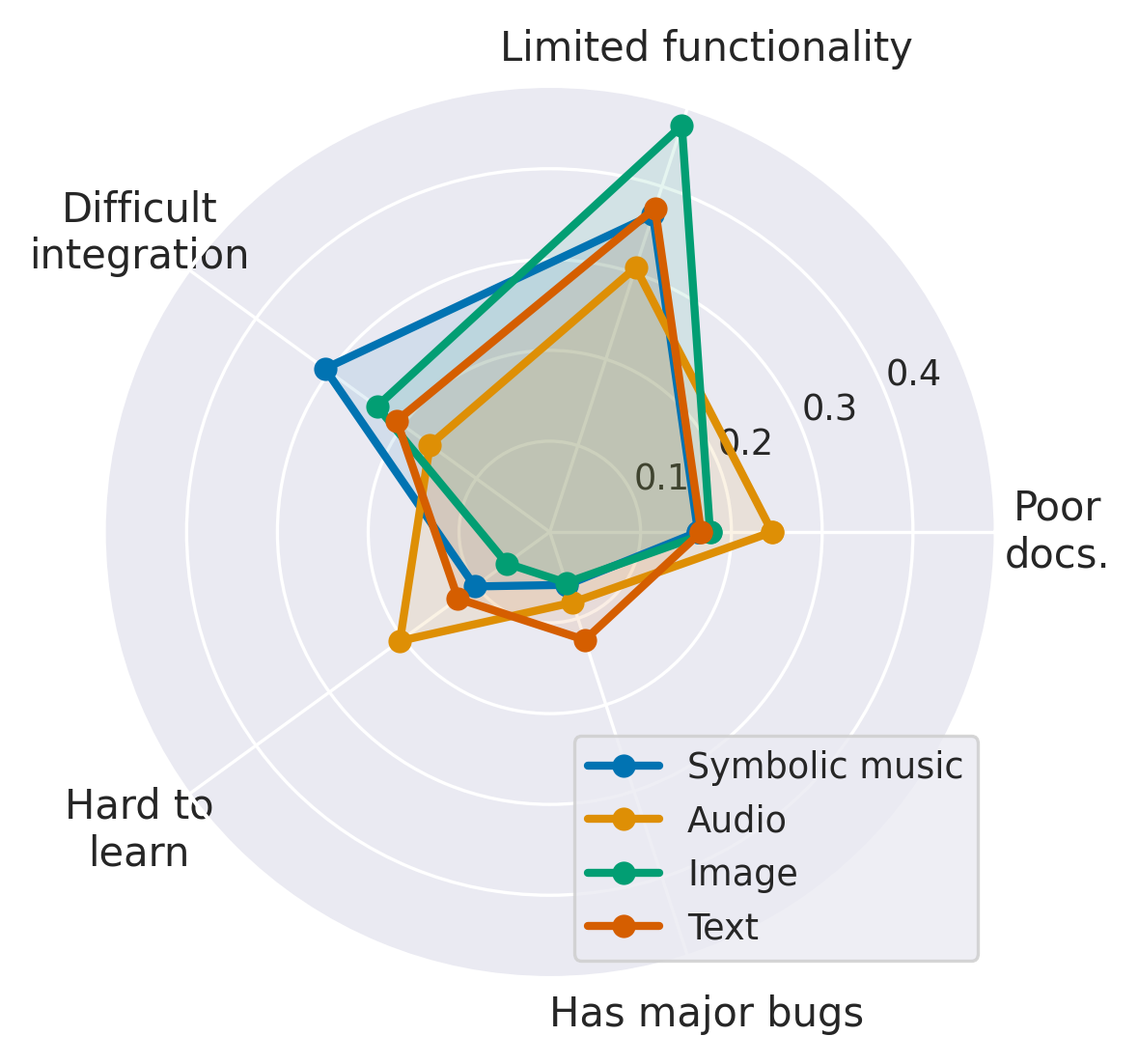}
    \caption{Percentage of reports per unmet need across modalities.}
    \label{fig:all-dislikes}
\end{figure}

\begin{figure*}[tb]
    \centering
    \includegraphics[width=\linewidth]{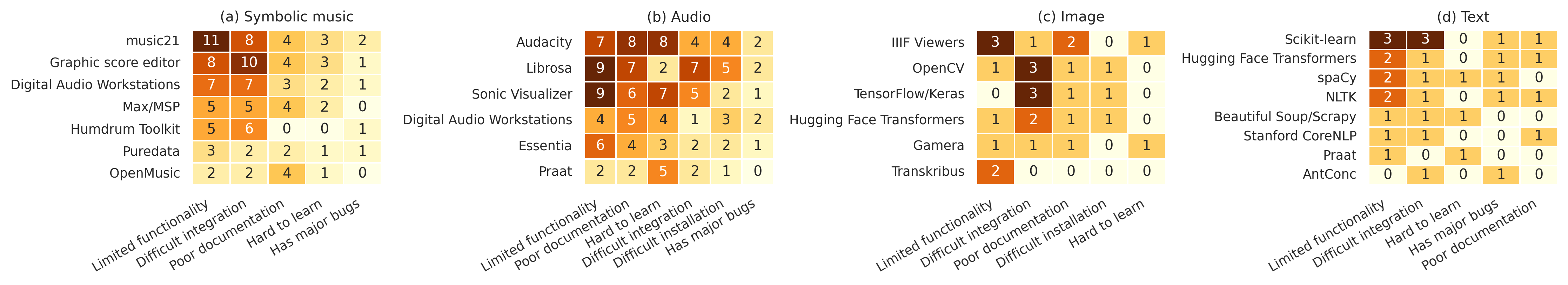}
    \caption{Tools vs. main concerns reported. Each value represents the number of participants who reported using a tool X (among all that they use) and also reported a concern Y (among all they reported). Values are filtered to show the top 75\% common matches within each domain.}
    \label{fig:tools_vs_disl}
\end{figure*}

As Figures~\ref{fig:all-dislikes}~and~\ref{fig:tools_vs_disl} show, limited functionality emerges as the consistent primary concern across all modalities, followed by integration difficulties and poor documentation. Notably, audio tools present a more significant learning curve challenge compared to other domains.

Regarding the importance of different factors when choosing a CM tool, the availability of well-documented APIs and examples emerged as one of the most critical considerations. This was followed by cross-platform compatibility and ease of integration with other software tools.

When asked about integration difficulties of data, methods, or tools, nearly half of the participants reported encountering such issues. Commonly mentioned challenges included the use of different standards across tools, limited support for various data formats, and insufficient capabilities for data translation.

\section{Discussion}\label{sec:discussion}

While our survey provides valuable insights into the CM community's practices and needs, certain limitations should be acknowledged. The sample size of 51 responses may not fully capture the global perspectives within the CM community. Additionally, the distribution through academic mailing lists and professional networks may introduce a bias toward more established researchers and those already engaged in computational approaches, potentially underrepresenting emerging practitioners. Despite these limitations, the consistent patterns observed across different modalities suggest that our findings can offer guidance for future tool development and community practices. 

Next, we will discuss the results presented in the previous section and try to extrapolate useful insights that can positively contribute to the CM community.

\paragraph{Aligning Tools with Practice}
This study represents, to our knowledge, the first systematic investigation of the practical needs of CM practitioners. The results provide a basis for computer scientists to align their development efforts with the requirements of the musicological community. Research of this nature is vital for effectively guiding multidisciplinary fields like CM towards successful outcomes. Priorities for future tool development should be informed by, or directly geared towards, addressing the most common tasks undertaken by musicologists, thus reducing the heavily reported limited functionality across tools. Historically, reciprocal needs and challenges between computer scientists and musicologists have often been addressed only within the confines of specific projects or research teams. Adopting a more participatory approach to tool development would optimize resources and reduce the redundant development of similar features across independent tools. \textbf{Take-away}: The CM community should attempt to establish tighter connections and tools for monitoring the overall mismatch among the different sides of the community.

\paragraph{Towards Integrated Music Analysis}
Difficulties in integrating tools and data consistently ranked among the top concerns reported by researchers. This finding is particularly significant given that over 85\% of surveyed users engage with multiple modalities in their research workflows. The necessity for cross-modal analysis exposes a key weakness in the current CM landscape: powerful tools often exist in isolation, impeding integrated analysis of music in its entirety. Future tool development should prioritize interoperability to facilitate seamless data sharing and synchronization between different modalities~\cite{ludovico2019adoption,LewisEtAl2018}. While this topic has been explored within MIR (e.g., audio-to-score alignment, lyrics alignment), practical tools for integrating information and analysis derived from disparate modalities remain uncommon~\cite{devaney2024pyampactscoreaudioalignmenttoolkit}, especially within user-friendly GUI environments. \textbf{Take-away}: A larger effort towards the standardization of CM tools and their interoperability and integration would help towards a more holistic analysis of music.

\paragraph{The Data Scale Challenge}
The analysis highlights the prevalence of small to medium-sized datasets across
various musical modalities (Figure~\ref{fig:corpora-size}). This presents a
notable challenge for CM, particularly in the context of leveraging Deep
Learning (DL) methods, which have become standard in other areas of MIR but are
less frequently applied in CM. This limitation is not solely technological; it
is often intrinsically linked to the nature of the data itself, such as the
scarcity or incompleteness of historical musical sources. Consequently,
researchers face fundamental hurdles in developing reproducible,
state-of-the-art DL models for musicological analysis. This necessitates the
exploration of innovative approaches capable of extracting meaningful insights
from limited data. For instance, strategies like Active Learning are being
explored to minimize the manual annotation effort required for Optical Music
Recognition (OMR) in historical manuscripts, directly tackling the data scarcity
problem \cite{sharma2025experimenting}. The infrequent adoption of advanced
NLP methods, such as transformer-based architectures, among musicologists
further underscores this point. \textbf{Take-away}: developing tools for CM
using LLM-based approaches may be geared towards different purposes than what
musicologists seek in the short term.

\paragraph{Bridging the Interface Gap}
A clear preference for GUIs among musicologists was observed, contrasting with the tendency of computer scientists to develop libraries and command-line tools. This divergence highlights a critical misalignment in development priorities. While the limited technical background of some humanists surveyed likely contributes to this preference, the increasing integration of LLMs into everyday tools may be perceived by some as diminishing the urgency for widespread technical training in humanities curricula~\cite{karjus2025machine}. However, rather than assuming musicologists will acquire extensive coding skills, a more pragmatic approach involves computer scientists developing software that facilitates interaction for non-technical users. Conversely, a valid concern exists that complex visual interfaces could potentially introduce bias into the analyses~\cite{miniukovich2018visual}. Providing modular building blocks in libraries and command-line tools offers developers a defensive position against this. The challenge lies in fostering both approaches: enabling technically proficient individuals to investigate potential biases and develop novel analysis methods while simultaneously providing accessible visual and interactive tools for musicologists with less technical expertise. \textbf{Take-away}: By creating or contributing to visual-based tools, or integrating them into command-line interfaces and libraries, developers would better serve the practical needs of the musicological community while reducing the technical barriers to adoption.

\section{Conclusions}\label{sec:conclusions}

The results reveal not only the primary concerns of practitioners in the field regarding the use of CM tools but also offer valuable insights into their workflows, tasks, tool preferences, and data practices. These findings shed light on how CM tools are integrated into daily research activities and point to specific areas where usability, functionality, and support can be improved.

Our survey highlights the significant gap that still exists between the music and developer communities, emphasizing that deeper collaboration is necessary to create tools that better serve the needs of computational musicologists. However, such collaboration is inherently challenging, primarily due to the mutual lack of domain-specific knowledge that prevents practitioners from articulating their precise requirements effectively. 

Consequently, targeted surveys and similar investigative strategies become crucial in bridging this communication divide and gaining a comprehensive understanding of the nuanced needs across these interdisciplinary domains.

\begin{acks}
This work has been funded by the European Union (Horizon Programme for Research and Innovation 2021-2027, ERC Advanced Grant “The Italian Lauda: Disseminating Poetry and Concepts Through Melody (12th-16th century)”, acronym LAUDARE, project no. 101054750). The views and opinions expressed are, however, only those of the author and do not necessarily reflect those of the European Union or the European Research Council. Neither the European Union nor the awarding authority can be held responsible for such matters.
\begin{figure}[H]
    \centering
    \includegraphics[width=0.65\linewidth]{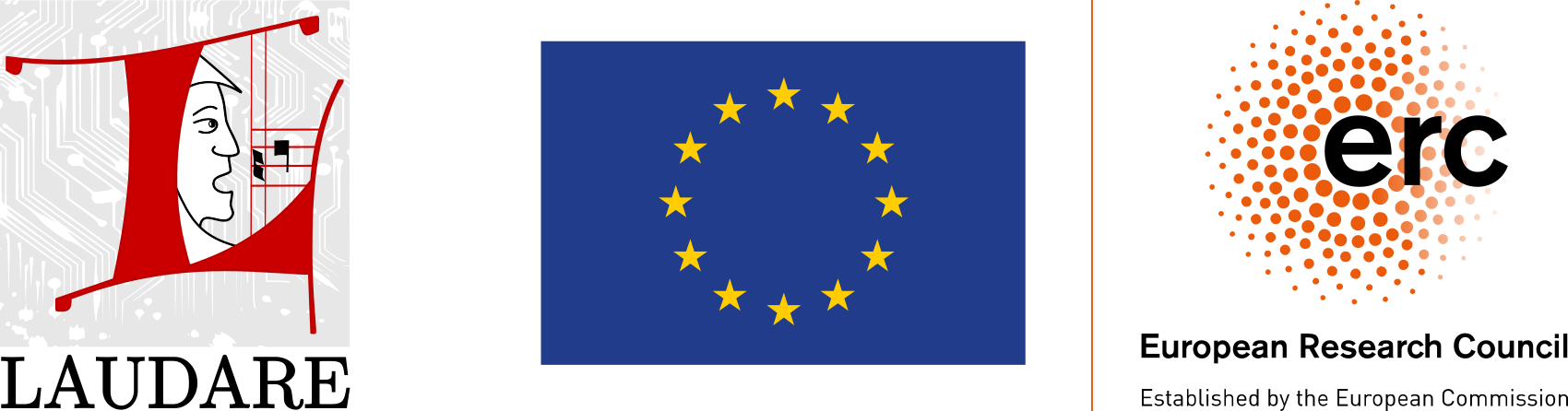}
\end{figure}
\end{acks}

\bibliographystyle{ACM-Reference-Format}
\bibliography{sample-base,related_works}

\end{document}